\documentclass[prl,twocolumn,nofootinbib, preprintnumbers, superscriptaddress]{revtex4}

\usepackage{float}
\usepackage{slashed}
\usepackage{amsmath,amssymb,slashed,braket}
\usepackage{graphicx}
\usepackage{epstopdf}
\usepackage{comment}
\usepackage{float,appendix}
\usepackage[colorlinks=true,
            linkcolor=blue,
            urlcolor=blue,
            citecolor=green,          
            bookmarks=true,
            bookmarksnumbered=true,
            breaklinks=true,
            pdfpagemode=Fullscreen,
            pdfstartview=FitBH]{hyperref}
\usepackage{esint}
\usepackage[normalem]{ulem}
\usepackage{color}

\definecolor{Orange}{cmyk}{0,0.61,0.87,0}
\definecolor{JungleGreen}{cmyk}{0.99,0,0.52,0}
\definecolor{OliveGreen}{cmyk}{0.64,0,0.95,0.40}
\definecolor{Brown}{cmyk}{0,0.81,1,0.60}
\definecolor{RoyalBlue}{cmyk}{0.71,0.53,0,0.12}
\definecolor{Gray}{cmyk}{0,0,0,0.40}
\definecolor{LightPink}{cmyk}{0.0,0.25,0,0}
\definecolor{LLightPink}{cmyk}{0.0,0.10,0,0}
\definecolor{LightBlue}{cmyk}{0.25,0,0,0}
\definecolor{LightGray}{cmyk}{0,0,0,0.2}

\usepackage{xcolor}
\definecolor{gesfpurple}{rgb}{0.47,0.19,0.42}

\definecolor{gesflanse}{rgb}{0.00,0.50,0.50}

\definecolor{gesfblue}{rgb}{0.08,0.42,0.76}

\definecolor{gesfred}{rgb}{1,0,0}

\definecolor{gesfwhite}{rgb}{1,1,1}

\definecolor{gesfblack}{rgb}{0,0,0}

\newcommand{\geqn}[1]{Eq.\,\hypersetup{linkcolor=blue}(\ref{#1})\hypersetup{linkcolor=blue}}
\newcommand{\gfig}[1]{{\hypersetup{linkcolor=violet}Fig.\,\ref{#1}\hypersetup{linkcolor=blue}}}
\newcommand{\gtab}[1]{{\hypersetup{linkcolor=gesflanse}Tab.\,\ref{#1}\hypersetup{linkcolor=blue}}}

\graphicspath{{Figures/}}

\begin{document}

\title{\Large Testable Inverse  Seesaw Motivated from a High Quality QCD Axion}

\author{Yannis Georis}
\email{yannis.georis@ipmu.jp}
\affiliation{Kavli IPMU (WPI), UTIAS, University of Tokyo, Kashiwa, 277-8583, Japan}

\author{Jie Sheng}
\email{jie.sheng@ipmu.jp}
\affiliation{Kavli IPMU (WPI), UTIAS, University of Tokyo, Kashiwa, 277-8583, Japan}

\author{Salvador Urrea}
\email{salvador.urrea@ijclab.in2p3.fr}
\affiliation{P\^ole Th\'eorie, Laboratoire de Physique des 2 Infinis Ir\`ene Joliot-Curie (UMR 9012), CNRS/IN2P3, 15 Rue Georges Clemenceau, 91400 Orsay, France}

\author{Tsutomu T. Yanagida}
\email{tsutomu.tyanagida@gmail.com}
\affiliation{Kavli IPMU (WPI), UTIAS, University of Tokyo, Kashiwa, 277-8583, Japan}
\affiliation{Tsung-Dao Lee Institute, Shanghai Jiao Tong University, 201210, China}


\begin{abstract}

The QCD axion remains one of the most compelling solutions to the strong CP problem. 
Meanwhile, the type-I seesaw mechanism offers an elegant explanation for the lightness of the observed neutrino masses; however, its extremely heavy Majorana states place it far beyond experimental reach. 
Low-scale alternatives such as the inverse seesaw improve testability but typically lack a strong theoretical motivation. 
In this paper we bridge this gap by showing that gauging the discrete symmetry $\mathbb Z_4 \times \mathbb Z_3$—motivated by the internal structure of the Standard Model—naturally yields a QCD axion with a high-quality Peccei–Quinn symmetry solving the strong CP problem, while simultaneously enforcing the field content and hierarchy required for a natural inverse seesaw. 
The resulting model is highly predictive and has the potential to be fully tested by future experiments. 
Beyond addressing the strong CP problem and the origin of neutrino masses, our scenario also contains a viable dark-matter candidate and offers potential mechanisms for generating the baryon asymmetry of the Universe.

\end{abstract}

\maketitle

\section{Introduction}

Since its establishment, the Standard Model (SM) of particle physics has achieved many remarkable successes, both theoretically and experimentally. 
However, to this day it remains incomplete. In particular, the SM cannot account for the existence of the (tiny) neutrino masses and oscillations~\cite{Esteban:2024eli}, as well as for the apparent preservation of the CP symmetry by strong interactions, the so-called strong CP problem~\cite{Peccei:1977hh,Peccei:1977ur,Wilczek:1977pj,Weinberg:1977ma}.

On the one hand, the type-I seesaw mechanism\footnote{The term \textit{seesaw mechanism} was first coined by one of the authors, T. T. Y., at a conference in 1981~\cite{INS1981}.}~\cite{Minkowski:1977sc,Yanagida:1979as,Yanagida:1979gs,Gell-Mann:1979vob} provides one of the most elegant explanation for the existence and smallness of the neutrino masses. 
Although the masses $M_N$ of the hypothetical right-handed (RH) Majorana neutrinos $N$ proposed in these initial scenarios were typically extremely large, $M_{N} \sim \mathcal{O}(10^{10-15})$ GeV, far beyond the reach of any current and foreseeable experiments, various low-scale alternatives have since been proposed, see e.g. \cite{Drewes:2013gca,Abdullahi:2022jlv} for reviews on the topic.
In these scenarios, the smallness of the neutrino masses is ensured by an approximate symmetry \cite{Shaposhnikov:2006nn,Kersten:2007vk,Moffat:2017feq} while remaining consistent with testable Yukawa couplings.
The so-called inverse seesaw mechanism (ISS)~\cite{Mohapatra:1986bd,Gonzalez-Garcia:1988okv,Gavela:2009cd,BERNABEU1987303,Mohapatra:1986aw} 
lies among the most popular realizations of this idea. 
By introducing a small lepton-number-violating parameter $\mu$, it can naturally generate the observed neutrino masses while keeping the RH neutrino masses at or below the TeV scale, well within the reach of current and future collider and intensity-frontier experiments. 
However, the presence of additional light Majorana fermions in the inverse seesaw mechanism is typically imposed by hand and lacks a compelling theoretical motivation.

On the other hand, the QCD axion, as the Nambu-Goldstone boson associated with the $U(1)$ Peccei-Quinn (PQ) symmetry, provides the most compelling solution to the strong CP problem by dynamically relaxing the CP-violating QCD angle $\theta$ to zero. 
However, this solution can be easily spoiled by the appearance of Planck-suppressed operators that break this global PQ symmetry and in turn modify the QCD axion potential to generate a non-vanishing value for $\theta$ above the experimental bound of $\mathcal{O}(10^{-10})$~\cite{Kamionkowski:1992mf,Holman:1992us}. 
Preserving the quality of the PQ symmetry requires an enormous fine-tuning at the level of $\sim 10^{-100}$ \cite{Kamionkowski:1992mf,Holman:1992us}. 
This severe fine-tuning issue is commonly referred to as the axion quality problem. 

In a recent study~\cite{Sheng:2025sou}, a subset of the authors have shown that the discrete symmetry $\mathbb Z_4 \times \mathbb Z_3$ embedded in the SM fermion structure not only leads to the emergence of a PQ symmetry in a natural way, but also predicts a high-quality QCD axion that resolves the strong CP problem. 
In this work, we first propose a bridge between this axion solution and the inversee seesaw mechanism. 
We point out that this framework naturally motivates light Majorana species, which can generate the small neutrino masses and predicts right-handed neutrinos with masses around the GeV scale. Such a scenario could be tested at a variety of near-future experiments~\cite{Aberle:2839677,Antel:2023hkf,Coloma:2020lgy,Coloma:2023adi,Blondel:2022qqo,FASER:2018eoc}. 
This minimal extension of the SM addresses multiple problems and leads to a rich phenomenology.

The paper is structured as follows. We start by reviewing the model first proposed in \cite{Sheng:2025sou}, emphasizing how the setup addresses the axion high-quality problem in a minimal way. 
The discussion then shifts to the neutrino sector, where the mechanism responsible for generating neutrino masses is examined. 
Constraints on the model parameter space set by various current and near-future experiments are analyzed next.  A brief discussion of the prospects for realizing leptogenesis within this framework is also included. The paper concludes with a summary and discussion of the main results.

\begin{table*}[t!]
    \centering
    \large
\begin{tabular}{p{2cm} p{1cm} p{1cm} p{1cm} p{1cm} p{1cm} p{1cm} p{1cm}}
\hline
Fields & $T$ & $\bar F$ & $\bar N$ & $\chi$ & $H_1$ & $H_2$ & $\Phi$ \\
\hline
\hline
\quad $\mathbb{Z}_4$ & 1 & 1 & 1 & 0 & -2 & -2 & -1 \\
\hline
\quad$\mathbb{Z}_3$ & 1 & 1 & 1 & -1 & ~2 & ~1 & -1\\
\hline
\end{tabular}
\caption{Charge assignments of the different fields under the $\mathbb Z_4$ and $\mathbb Z_3$ discrete gauge symmetries. 
}
    \label{tab:chargeassignments}
\end{table*}

\section{Brief review on the high quality QCD axion in the Standard Model}

The SM of particle physics is built upon the gauge symmetry $SU(3)_c \times SU(2)_L\times U(1)_Y$. In addition, certain gauged discrete symmetries are naturally embedded in the  structure of the SM fermions. A well-known example is the $\mathbb Z_4$
symmetry, which is anomaly-free with respect to all SM gauge interactions if all chiral fermions carry a charge of $+1$ under it.
However, when non-perturbative effects are taken into account, this symmetry suffers from the so-called \textit{Dai–Freed anomaly} \cite{Dai:1994kq,Yonekura:2016wuc,Garcia-Etxebarria:2018ajm}, which can be canceled by introducing three right-handed neutrinos $\bar N_i$. 
Similarly, since the fermion sector consists of three generations, it naturally accommodates a $\mathbb Z_3$ gauged discrete symmetry, which is anomaly-free when all chiral fermions carry charge $+1$. 
The corresponding Dai–Freed anomaly in this case can also be canceled by introducing three generations of Majorana singlet field $\chi_i$ with $\mathbb Z_3$ charge $-1$ \cite{Hsieh:2018ifc}. 
Therefore, the gauge group of the SM can be extended by the gauge discrete symmetries $\mathbb Z_4 \times \mathbb Z_3$.

To illustrate this model in a simple way, the quarks, leptons, and Higgs doublets are organized into $SU(5)$ representations:
\begin{subequations}
\begin{align}
    T ({\bf 10}) & \equiv \{q, \bar{u}, \bar{e} \} , \\
    \bar{F} ({\bf 5}^*) & \equiv \{\bar{d}, \ell \},  \\
   H({\bf 5}^*) &\equiv \{H_1, H_2\}.
\end{align}
\end{subequations}
However, we emphasize that this is done merely for organizational purposes, and we do not assume any grand unification of the SM gauge groups, nor do we impose any GUT relations among the gauge and Yukawa couplings.
The minimal charge assignment of the various fermionic fields and two Higgs doublets under the discrete gauge symmetries is summarized in \gtab{tab:chargeassignments}

We find that this minimal extension of SM naturally accommodates a high-quality QCD axion, providing a solution to the strong CP problem \cite{Sheng:2025sou}. Due to the presence of the $\mathbb Z_3$ symmetry, two Higgs doublets $H_1, H_2$ are required in order to write all SM fermion mass terms, 
$\mathcal{L}_{\text{Y}}^{\text{SM}} \sim T TH_1^\dagger/2 + T \bar{F} H_2 + 
\bar N\bar FH_1^\dagger  + \text{h.c.}$. This naturally introduces the following additional global $U(1)$ symmetry, identified as the PQ symmetry \cite{Peccei:1977hh,Peccei:1977ur}: 
\begin{subequations}
\begin{align}
    \{\ell, q\} &\rightarrow e^{-i \alpha}\{\ell, q\}, \\
    \{\bar e, \bar u, \bar d, \bar N\} &\rightarrow e^{-i \alpha}\{\bar e, \bar u, \bar d, \bar N\},\\
    \{H_1, H_2\} &\rightarrow \{e^{-2i \alpha}H_1, e^{+2i \alpha}H_2\},\\
    \{\Phi\} &\rightarrow \{e^{+i \alpha}\Phi\}, \\
    \{\chi\} &\rightarrow \{e^{-2i \alpha}\chi\}.
\end{align}
\label{PQcharges}
\end{subequations}

To generate mass terms for the right-handed neutrino $N$ and field $\chi$, an additional scalar field $\Phi$ is introduced,
which also serves as the PQ field. Assuming the charges shown in \gtab{tab:chargeassignments}, the higher-dimensional operator $H_1^\dagger H_2 \Phi^{*4}/ M_{\text{Pl}}^2$, suppressed by powers of the (reduced) Planck mass $M_{\text{Pl}} \simeq 2.4 \times 10^{18}\,$GeV, 
is also allowed under the two discrete symmetries. 
The PQ symmetry breaking happens when the field $\Phi$ gains its vacuum expectation value $\braket{\Phi} \equiv F_a$. The associated pseudoscalar Nambu-Glodstone boson, axion $a$, then provides a dynamical solution to the strong CP problem \cite{Wilczek:1977pj,Weinberg:1977ma}.

However, the axion solution typically suffers from a severe fine-tuning issue, known as the axion quality problem.
Quantum gravity effects are believed to violate explicitly all global symmetries, including the PQ symmetry, by introducing effective higher-dimensional potential operators
$V_g (\Phi) \sim \Phi^n/M_{\text{Pl}}^{n-4}$ \cite{Giddings:1988cx,Coleman:1988tj,Gilbert:1989nq,Hawking:1975vcx}.
Typically, operators with small dimension $n$ spoil the QCD axion potential, and suppressing them requires a severe fine-tuning of $\mathcal{O}(10^{-100})$ of the couplings \cite{Kamionkowski:1992mf,Holman:1992us}.

In our model, the presence of a discrete gauge symmetry $\mathbb Z_4 \times \mathbb Z_3$, which cannot be broken by gravity, restricts the lowest-dimensional gravity-induced operator to the form $V_g \sim g \Phi^{12}/M_{\text{Pl}}^8$. This operator is naturally suppressed by eight powers of the Planck scale, allowing the axion to remain of high quality as long as $F_a \lesssim 2 \times 10^{12}~\mathrm{GeV}$, without fine-tuning of the coupling $g$~\cite{Sheng:2025sou}.

In addition, the misalignment mechanism~\cite{Preskill:1982cy,Abbott:1982af,Dine:1982ah} naturally produces axion dark matter in the early Universe. Requiring the axion to constitute the dominant component of dark matter while keeping the initial misalignment angle of order one implies $F_a \gtrsim 10^{11}~\mathrm{GeV}$. 
Therefore, this scenario yields a definite prediction for the axion dark matter mass:
\begin{equation}
3 \times 10^{-5}~\mathrm{eV} \lesssim m_a \lesssim 5 \times 10^{-4}~\mathrm{eV},
\end{equation}
which falls within the reach of upcoming haloscope searches~\cite{AxionLimits}.

Furthermore, the anomaly-free condition of the $\mathbb Z_3$ symmetry predicts the existence of three light fields $\chi_i$.
Their presence lead the model to embed a testable inverse seesaw mechanism, providing another potential smoking gun, which we discuss below as the main point of this paper.

\section{Neutrino Mass generation From Inverse Seesaw}

In this section, we describe the neutrino sector and show that neutrino masses and their smallness arise as a natural prediction of the inverse seesaw mechanism embedded in the $\mathbb Z_4 \times \mathbb Z_3$ symmetry extension of the SM. The particle content and symmetries of our model allow us to write the following mass and mixing terms for the right-handed neutrino $N$ and Majorana particle $\chi$:
\begin{equation}
\begin{aligned}
\mathcal{L}_{\nu} \supset\;&
y_{\nu}\,\bar{N}\,\ell\,H_1^\dagger
+\frac{1}{2}\frac{c_{N}}{M_{\rm Pl}}\Phi^2\,\bar{N} \bar{N} \\
&+\frac{1}{3!}\frac{c_M}{M_{\rm Pl}^2}{\Phi^*}^3\,\bar{N} \chi
+\frac{1}{2}\frac{1}{4!}\frac{c_\mu}{M_{\rm Pl}^3}\Phi^4\,\chi \chi
+\text{h.c.} \ ,
\end{aligned}
\label{eq:Lnu}
\end{equation}
where $c_M$ is a $3\times3$ complex matrix and $c_{\mu}$, $c_{N}$ are $3\times3$ symmetric matrices.\footnote{Here we adopt the convention for Weyl spinor contractions described in Ref.~\cite{Schwartz:2014sze}. For example, we define $\chi \chi \equiv \chi_\alpha \chi^\alpha \equiv \chi_\alpha \epsilon^{\alpha \beta} \chi_\beta$, with $\epsilon^{\alpha \beta}$ being the totally antisymmetric $2\times 2$ tensor.} 
From a simple power counting, it is apparent that the Majorana mass term of $N$ should, in principle, constitute the dominant contribution to the light neutrino masses.
Although neutrino mass generation works perfectly well in this setup, this would severely limit the allowed range of RH neutrino Yukawa coupling and masses.
Even an $N$ mass as low as $\sim 10^6\,$GeV remains untestable in direct detection experiments.
Therefore, we take $c_N$ to be negligibly small, which is technically natural~\cite{tHooft:1979rat} since a 
$\mathbb{Z}_7$ symmetry\footnote{The mentioned $\mathbb{Z}_7$ symmetry is defined by the following charge
assignments: $T(3)$, $\bar{F}(1)$, $\bar{N}(5)$, $\chi(5)$, $H_1(6)$, $H_2(3)$, and $\Phi(1)$.
} emerges in the limit of $c_N \rightarrow 0$.
The ${\Phi^*}^3 \bar N \chi$ term violates the PQ symmetry and, at one loop, induces a contribution to the effective potential $\propto \Phi^{12}$ that can potentially endanger the axion quality. 
However, this contribution scales with $M_N$, and the axion quality can therefore still be protected if the Majorana mass terms for $\bar N$ are suppressed by a small value of $c_N$. 
This observation even supports the present inverse seesaw mechanism.\footnote{In Ref.~\cite{Sheng:2025sou}, where this model was initially introduced, an additional $\mathbb{Z}_2$ gauge symmetry was assumed to suppress the $c_M$ term.}

After the PQ and EW symmetry breakings, the neutrino mass matrix in the $(\nu_L, N, \chi)$ basis takes the form
\begin{equation}
\label{eq:mass_matrix} \mathcal{M}_\nu=\begin{pmatrix} 0 & m_D & 0\\ m_D^T & 0 & M\\ 0 & M^T & \mu \end{pmatrix} \ ,
\end{equation}
where $m_D$ is the standard Dirac mass, $M$ represents the mixing between $N$ and $\chi$, and $\mu$ is the Majorana mass term for $\chi$. 
They are defined as
\begin{subequations}
\begin{align}
   m_D &\equiv y_\nu \frac{v}{\sqrt{2}}, \\
   M &\equiv \frac{1}{3!}\frac{c_M}{M_{\rm Pl}^2}(F_a)^3 , \\
   \mu &\equiv \frac{1}{4!}\frac{c_\mu}{M_{\rm Pl}^3}(F_a)^4. 
\end{align}
\label{masses}
\end{subequations}
Here, we take $c_N = 0$ and $v$ to be the electroweak vacuum expectation value\footnote{For simplicity we haven taken $v_1=v_2=v/\sqrt{2}$. This, however, does not impact our results, as any other choice would merely correspond to a rescaling of the Yukawa couplings.
} (vev).
The mass matrix structure of \geqn{eq:mass_matrix} corresponds to the inverse seesaw mechanism. In our specific case, it realizes ISS(3,3), where three $(N,\chi)$ pairs are introduced.
This configuration represents the third-minimal implementation of the inverse seesaw capable of successfully reproducing neutrino oscillation data~\cite{Abada:2014vea}.

\begin{center}
    \begin{figure*}[t!]
        \centering     \includegraphics[width=0.92\textwidth]{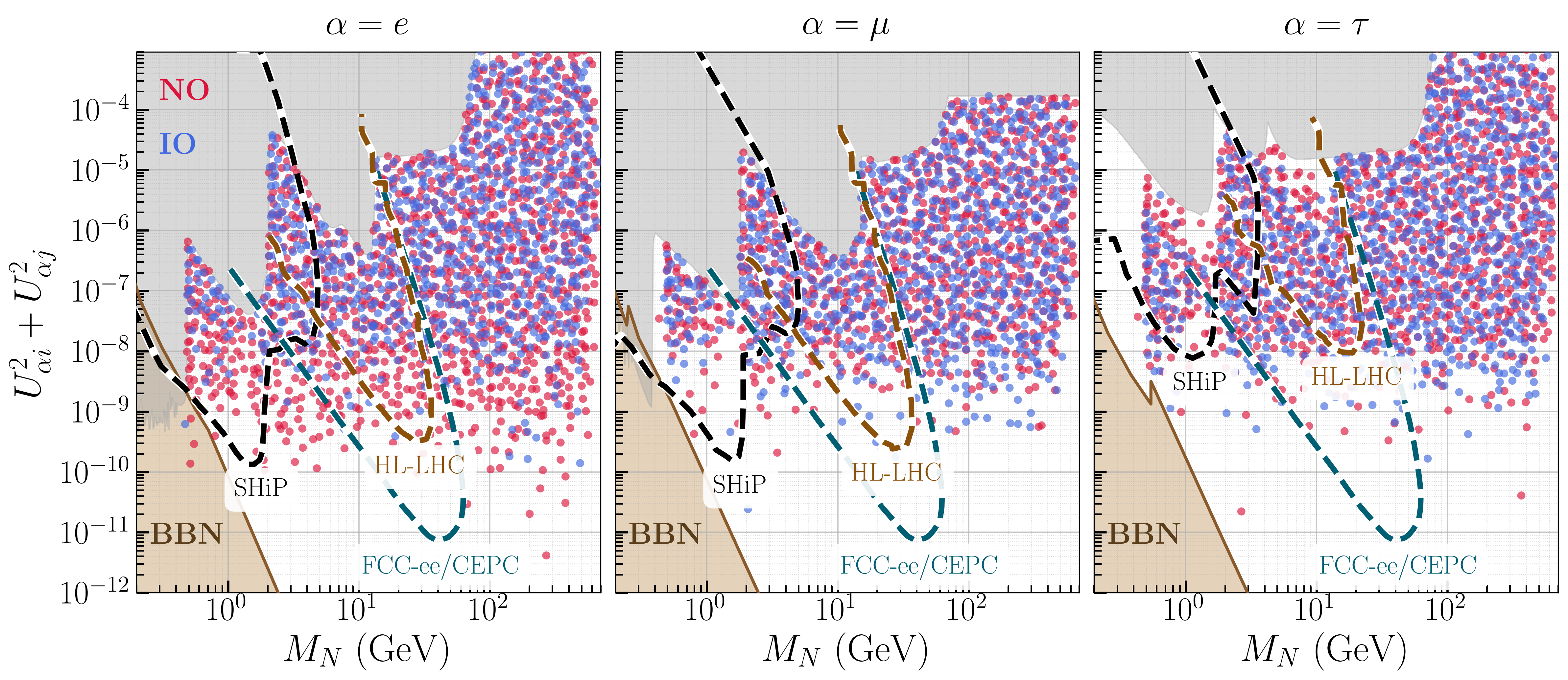}
        \caption{
        Points consistent with neutrino oscillation data and allowed by our model. Red (blue) points correspond to normal (inverted) mass ordering. The gray shaded regions denote current experimental constraints~\cite{Fernandez-Martinez:2023phj}, and the brown shaded area is disfavored by BBN bounds extracted from~\cite{Abdullahi:2022jlv,Boyarsky:2020dzc}. The dashed lines indicate the projected sensitivities of the SHiP experiment \cite{Aberle:2839677,Antel:2023hkf}, the HL-LHC (CMS displaced-vertex searches) \cite{Izaguirre:2015pga,Drewes:2019fou} and FCC-ee/CEPC \cite{Blondel:2022qqo,CEPCStudyGroup:2018ghi,Ai:2025cpj}. These sensitivity lines have been derived in a simplified framework assuming a unique heavy neutrino $N$ solely coupled to the SM flavor $\alpha$.}  \label{fig:HNL}
    \end{figure*}
\end{center}

The effective light neutrino mass matrix in the ISS mechanism takes the form
\begin{equation}\label{eq:mass_light}
\mathcal{M}^{\text{light}}_{\nu} \simeq m_D^T M^{-1}\mu(M^{-1})^T m_D.
\end{equation}
Unlike the conventional seesaw mechanism, where the smallness of neutrino masses arises from the heaviness of the right-handed neutrinos, in the ISS the light neutrino masses are suppressed by the small Majorana masses $\mu \ll M$ of $\chi$. In our model, both $M$ and $\mu$ are generated from Planck-suppressed operators involving the PQ field, and therefore inherit a characteristic scaling with the axion decay constant $F_a$. For $F_a$ close to its maximal value $2\times 10^{12}\,$GeV, one naturally obtains $\mu \lesssim 280\,$keV and a hierarchy of about $6\text{--}7$ orders of magnitude between $\mu$ and $M$. This predicts the light-neutrino masses scale as 
\begin{equation}
    m_\nu \simeq \frac{3}{2}\,\frac{c_\mu}{c_M^2}\,\frac{M_{\rm Pl}}{F_a^2}\,m_D^2,
\end{equation}
which makes the smallness of $m_\nu$ directly correlated with the high quality of the axion. As a concrete illustration, taking $y_{\nu} \simeq y_e$ so that $m_D \simeq 0.5~\mathrm{MeV}$ and considering $F_a = 2\times 10^{12}~\mathrm{GeV}$ with $c_\mu/c_M^2 \sim 1$ leads to a prediction $m_\nu \simeq 0.1~\mathrm{eV}$, which is consistent with the current cosmological upper bound on the sum of neutrino masses~\cite{DESI:2024mwx,eBOSS:2020yzd,Palanque-Delabrouille:2019iyz,Naredo-Tuero:2024sgf,Craig:2024tky}.

Neglecting for a moment the mixing with the light neutrinos, the heavy neutrino mass matrix reads
\begin{equation} 
\mathcal{M}_H\simeq\begin{pmatrix} 0 & M \\ M & \mu \end{pmatrix}\ .
\end{equation}
We can then diagonalize this six-by-six matrix to obtain the heavy neutrino masses
\begin{equation}\label{eq:heavy_masses}
M_N^{i,\pm} \simeq M_i \pm \frac{\mu_i}{2},
\end{equation}
where $M_i$ and $\mu_i$ ($i = 1,2,3$) are the eigenvalues of the $M$ and $\mu$ matrices, respectively. 
Overall, for the parameter range specified in \geqn{eq:scan}, this model therefore typically predicts the existence of three pseudo-Dirac pairs with masses\footnote{The lower bound is not a direct prediction of the model and arises instead from the strong constraints from BBN and direct detection experiments on the existence of feebly interacting particles below this mass.}
\begin{align}
\label{eq:typicalmassHNL}
      0.1~\mathrm{GeV} \lesssim M_N^i\lesssim 600~\mathrm{GeV}\ ,
\end{align}
and tiny mass splittings, controlled by the lepton-number-violating parameters $\mu_i$, between the two chiral fermions within the pseudo-Dirac pair.
We note in passing that additional contributions to the heavy neutrino mass splittings at $\mathcal{O}(m_D^2/M)$ appear due to their mixing with left-handed neutrinos.

\section{Experimental Constraints and Future Projections}

The diagonalization of the neutrino mass matrix~\eqref{eq:mass_matrix} is done by the full unitary leptonic mixing matrix linking the flavor basis to the mass basis $\mathcal{U}^\dagger \mathcal{M}_\nu \mathcal{U}^* \equiv \text{diag} (m_i \cdots)$, with $m_i$ the light neutrino masses, and $\mathcal U$ can be conveniently expressed in block form as
\begin{equation}\label{eq:mixing_matrix}
\mathcal{U}=\left(\begin{array}{ll}
\mathcal{N} & \Theta \\
\mathcal{R} & \mathcal{S}
\end{array}\right),
\end{equation}
where $\mathcal{N}$ denotes the $3\times3$ effective PMNS matrix~\cite{Pontecorvo:1967fh,Maki:1962mu}, which determines the standard neutrino oscillations. The $3\times 6$ blocks $\Theta$ and $\mathcal{R}$ encode the active–sterile mixing entries that govern experimental signatures of heavy neutrinos, whereas the block $\mathcal S$ describes the mixing within the sterile sector and is free from observation.

Although the fields $N$ and $\chi$ are sterile under the SM gauge interactions, their mixing $\Theta$ with the active neutrinos induces interactions with the electroweak gauge bosons, through which they can be produced at various laboratory experiments. The associated Lagrangian reads 
\begin{align}
\label{eq:weak_WW}
\mathcal{L} \supset{} &
- \frac{g}{\sqrt{2}}\,\overline{N_i}\,\Theta^*_{\alpha i}\,\gamma^\mu e_{L\,\alpha}\,W^+_\mu \notag\\
&- \frac{g}{2\cos\theta_w}\,\overline{N_i}\,\Theta^*_{\alpha i}\,\gamma^\mu \nu_{L\,\alpha}\,Z_\mu + \text{h.c.}\ ,
\end{align}
with $g$ the weak coupling and $\theta_w$ the weak mixing angle. 
Because event numbers at accelerator-based experiments are typically sensitive to the absolute value squared of the mixing angle $\Theta_{\alpha i}$, it is convenient to define the following quantities
\begin{align}
    U_{\alpha i}^2 \equiv |\Theta_{\alpha i}|^2 \mbox{ and } U^2 \equiv  \sum_{\alpha} \sum_i U_{\alpha i}^2 \ .
\end{align}
In the seesaw limit $m_D \ll M$, we have $\Theta \sim m_D/M$ and the total mixing angle
squared scales as $U^2\sim (m_D/M)^2$. 
In particular, neglecting for a moment the matrix structure of these equations, the inverse seesaw relation~\eqref{eq:mass_light} implies that
\begin{equation}
U^2 \sim \frac{m_\nu}{\mu},
\label{theta}
\end{equation}
leading to interactions that are suppressed with respect to those of SM neutrinos, yet still potentially large enough to be testable in current and future experiments. We note in passing that, given that $\mu \ll M$ in the relevant parameter space, the total mixing angle is typically much larger than that of the normal seesaw case, $U^2\sim m_\nu/M$  ~\cite{Drewes:2019mhg}. 

To investigate the constraints and the detectability of the mixing angles, we employ the Casas–Ibarra parametrization~\cite{Casas:2001sr,Mukherjee:2022fjm} to perform a comprehensive scan over the parameter space of our model, with the resulting distributions shown in Figure~\ref{fig:HNL}. 
In this scan we fix $v=246~\text{GeV}$~\cite{ParticleDataGroup:2024cfk} and we vary the fundamental parameters of our model $F_a, c_M, c_{\mu}, y_{\nu}$ in the following ranges:
\begin{subequations}
\begin{align}
   10^{11} ~\mathrm{GeV} \lesssim F_a &\lesssim 2 \times 10^{12}~\mathrm{GeV} , \\
   0 \lesssim c_{M} &\lesssim (4\pi)^3 , \\
   0 \lesssim c_{\mu} &\lesssim (4\pi)^4. 
   \\
   0 \lesssim y_{\nu} &\lesssim 4\pi.
\end{align}
\label{eq:scan}
\end{subequations}
The upper limit on $F_a$ ensures a high-quality axion without requiring fine-tuning, while the lower limit guarantees that the axion can serve as a viable dark matter candidate, as discussed in the previous section. For the ranges of $c_M$ and $c_\mu$, we follow the naive dimensional analysis (NDA) approach~\cite{Manohar:1983md,Cohen:1997rt,Luty:1997fk,Georgi:1986kr}.\footnote{Note that in the previous section we assumed $\mathcal{O}(1)$ couplings for the operator $\Phi^{12}/M_{\text{Pl}}^8$ instead of applying NDA. This is because the validity of NDA for operators of multi external fields is questionable~\cite{Nishio:2012sk}, and we therefore adopt a more conservative assumption. In addition, since the operator has dimension $12$, the dependence of the decay constant $F_a$ on the coupling $g$ is only $F_a \propto |g|^{1/12}$. Consequently, even if 
$g$ varies by three orders of magnitude, 
$F_a$ changes only by a factor of order two.} 
For each parameter point, we impose consistency with the global fit to neutrino oscillation data~\cite{Esteban:2024eli}, as well as with the recent improvements in the determination of $\theta_{12}$ and $\Delta m^2_{21}$ by JUNO~\cite{JUNO:2025gmd}, at the $3\sigma$ confidence level.\footnote{The recent JUNO results have only a minor impact on our results.} 
This is, the block $\mathcal{N}$ in Eq.~\eqref{eq:mixing_matrix} must reproduce the observed PMNS matrix within $3\sigma$, while also yielding the correct mass-squared differences $\Delta m^2_{21}$ and $\left|\Delta m^2_{3l}\right|$. Satisfying neutrino data forces most of our viable points to lie in the range $10^{-5} < y_\nu < 10^{-1}$ or, equivalently, $1\,\text{MeV} \lesssim m_D \lesssim 20\,\text{GeV}$, with values comparable to those of other SM fermions.

Each parameter point corresponds to three pseudo-Dirac pairs $(i,j) = (4,5),(6,7),(8,9)$, and we assign to it three values of $U_{\alpha i}^2 + U_{\alpha j}^2$, which are then plotted as a function of $M_N$. We display only the point associated with the pair that has the largest active--sterile mixing, as these states would typically be the first accessible in current and upcoming experimental searches.

Even though we allow $c_M$ and $c_\mu$ to take small values in our scan, such values are already excluded by current constraints, since they lead to smaller $M$ or $\mu$ and to large $U_{\alpha i}^2$ according to \geqn{theta}, which are ruled out by existing experimental limits. For heavy-neutrino masses in the range $\mathcal{O}(0.1\text{--}1)\,\mathrm{GeV}$, these mixings allow for their copious production in charged and neutral meson decays (such as $\pi^\pm \to \ell^\pm N$, $K^\pm \to \ell^\pm N$, $D_{(s)}^\pm \to \ell^\pm N$ and $B^\pm \to \ell^\pm N$), and their subsequent detection through semileptonic channels (e.g. $N \to \ell_\alpha^- \pi^+$) or fully leptonic modes (e.g. $N \to \ell_\alpha^- \ell_\beta^+ \nu_\beta$), making fixed-target and beam-dump facilities particularly sensitive probes in this mass regime. At larger masses, production from electroweak bosons becomes more relevant (e.g. $W^\pm \to \ell^\pm N$ and $Z \to \nu N$), enabling searches at high-energy colliders where the characteristic displaced-vertex signatures of long-lived heavy neutrinos provide a powerful handle to suppress backgrounds~\cite{Antel:2023hkf,Beacham:2019nyx}.
To avoid overcrowding the figure, we retain only the points that are not currently excluded, or that lie very close to the present experimental limits, which are shown as shaded grey regions. 

We also show in brown the region disfavored by Big Bang Nucleosynthesis (BBN). In this region the heavy neutrinos would decay too late, with lifetimes roughly scaling as $\tau_N \sim 1/(U^2 M_N^5)$, leading to an excessive injection of energy into the primordial plasma and spoiling the successful predictions of light element abundances~\cite{Abdullahi:2022jlv,Boyarsky:2020dzc,Sabti:2020yrt}. 

In the mass range $M_N \sim (0.2\text{--}80)\,\mathrm{GeV}$, the majority of viable points in our model predict active–sterile mixings of $U^2 \gtrsim 10^{-9}$. 
This trend originates from the structure of Eq.~\eqref{theta}: the maximum value of $c_\mu$ enforces a lower bound on the total active–sterile mixing $U^2$, while the maximum value of $c_M$ bounds from above the heavy-neutrino masses $M_N$. 
It is important to stress that the true lower bound applies only to the total mixing $U^2$, not to the flavored quantities we show. 
The panels in Fig.~\ref{fig:HNL} display, for each parameter point, the largest mixing for a fixed flavor $\alpha$ among the three pseudo-Dirac pairs. 
One can suppress the mixing in a particular flavor by distributing $U^2$ very unevenly among flavors, i.e. $U_\alpha^2 / U^2 \ll 1$. 
Such strongly hierarchical flavor structures are statistically disfavored in a random scan of Yukawa matrices, so parameter points with extremely small $U_{\alpha i}^2 + U_{\alpha j}^2$ become increasingly rare. 
This causes the apparent lower boundary in the plots: it reflects the typical flavor pattern of viable points rather than a physical limit. A genuine lower bound applies exclusively to $U^2$, which cannot fall below $\sim 10^{-7}$. 
Therefore, a measurement of the mixings across all flavors down to the level of $\sim 10^{-7}$ would be sufficient to fully probe our model.

Moreover, Fig.~\ref{fig:HNL} shows both normal (NO) and inverted (IO) light neutrino mass orderings, distinguished by using different colors. In the IO case one has $m_1 \simeq m_2 \gg m_3$, so two light neutrinos are much heavier than the third, whereas in the NO case $m_3 \gg m_2 \simeq m_1$. Consequently, through Eq.~\eqref{theta} there is already a tendency to obtain larger $U_{\alpha i}^2$ for IO. Since our model contains three pseudo-Dirac pairs, it has enough freedom to make this difference negligible in the $\mu$ and $\tau$ sectors. However, it remains clearly visible in the leftmost panel of Fig.~\ref{fig:HNL} for $\alpha = e$, because the electron row of the PMNS matrix is the most asymmetric and hierarchical one, due to the smallness of $\theta_{13}$. As a result, the constraints coming from reproducing the observed PMNS structure are especially stringent in the electron sector.

This parameter region is precisely within the projected sensitivities of both future high-intensity fixed-target experiments such as SHiP~\cite{SHiP:2015vad} and displaced-vertex searches at future colliders, including the High-Luminosity LHC~\cite{ZurbanoFernandez:2020cco} and proposed facilities like FCC-ee~\cite{FCC:2018evy} and CEPC \cite{CEPCStudyGroup:2018ghi}. 
These prospects make our model particularly attractive from a phenomenological point of view, as a substantial portion of the parameter space can be probed in the coming decades.

Finally, we conclude this section by briefly commenting on the possibility to generate the baryon asymmetry of our universe within this scenario. 
Given the constraint in Eq.~\eqref{eq:typicalmassHNL} on the masses of the additional singlets $N$ and $\chi$, both leptogenesis from neutrino oscillations \cite{Akhmedov:1998qx,Asaka:2005pn} and resonant leptogenesis \cite{Pilaftsis:1997jf,Pilaftsis:2003gt,Pilaftsis:2005rv} could in principle contribute to the overall asymmetry generation.
Recent studies~\cite{Klaric:2020phc,Klaric:2021cpi} have established that the range of temperatures at which these mechanisms are effective overlap significantly in a minimal scenario with only two right-handed neutrino generations.
In addition, these studies have revealed a sizable parameter space consistent with the observed baryon asymmetry of our universe, which shows overlap with our \gfig{fig:HNL} in the low-mass regime ($U^2 \gtrsim 10^{-9}$ for $M_N \lesssim 100\,$GeV). 
The prospects for leptogenesis become even more favorable in scenarios with more than two fermionic singlet species, see e.g.~\cite{Abada:2018oly,Drewes:2021nqr,Drewes:2024pad}.  

However, this conclusion cannot be directly applied to the pure inverse seesaw mechanism for which generating the correct baryon asymmetry is more difficult~\cite{Abada:2015rta,Abada:2017ieq,Dolan:2018qpy}.
This observation stems from the fact that a unique parameter $\mu$ controls the amount of lepton number violation as well as the size of the heavy neutrino mass splittings and, indirectly, the rate of washout processes.
In the regime where it is resonantly enhanced, the generated asymmetry tends to be severely washed out at the same time.
We however note that the results of these previous studies do not automatically rule out our scenario as the origin of the matter-antimatter asymmetry as they typically rely on simplifying assumptions (e.g. large mass splittings between the pseudo-Dirac pairs) to avoid solving the complete set of (numerically stiff) quantum kinetic equations describing the RH neutrino time evolution.
Our scenario instead features three lepton number violating parameters $\mu_i$, and all three pseudo-Dirac pairs can have comparable masses and participate in leptogenesis.
Moreover, we note that this difficulty could also be alleviated by allowing for a (tiny) coupling between the light neutrinos and the singlet $\chi$.
This would provide an additional source of lepton number violation, whereas the typical mass splitting $\mu/M \simeq \mathcal{O}(10^{-7}-10^{-6})$ between the fermions constituting the pseudo-Dirac pair is compatible with leptogenesis in models with only $2$ singlet fermions~\cite{Klaric:2021cpi}. 
Furthermore, we have more parameters if we include a non-vanishing but small $c_N$, which generates the small Majorana masses for $N_i$.
Because of the extended field content and large parameter space of this model, a detailed numerical exploration of baryogenesis in the present scenario lies however beyond the scope of this paper. 
We note in passing that electroweak baryogenesis does not work in our scenario as no sufficient CP violation is introduced in the Higgs sector.

\section{Conclusion and Discussions}

The SM has two instanton solutions, one is the QCD instanton and the other the electroweak instanton. The QCD instanton has twelve quark and anti-quark zero modes and interestingly, the electroweak instanton has also twelve doublet quark and lepton zero modes \cite{Nomura:2000yk}. 
Thus, the SM exhibits a discrete $\mathbb{Z}_{12}$ symmetry, which is nothing but the $\mathbb{Z}_{4}\times \mathbb{Z}_{3}$ considered in this paper.

Therefore, a natural step toward beyond-the-SM physics is to gauge this discrete $\mathbb{Z}_4\times \mathbb{Z}_3$ symmetry. 
Under this symmetry, the SM alone would exhibit Dai-Freed gauge anomalies. 
However, these anomalies are canceled out if one introduces three right-handed neutrinos $\bar N_i$ and three chiral fermions $\chi_i$. 
Surprisingly, this model automatically has the PQ symmetry and the quality of the symmetry is sufficiently high to solve the strong CP problem~\cite{Sheng:2025sou}. 
Furthermore, the predicted three new fermions $\chi_i$ are precisely those required in the inverse seesaw mechanism, forming three pseudo-Dirac pairs together with the right-handed neutrinos $N_i$.

As highlighted in \geqn{eq:Lnu}, this scenario contains Planck-suppressed operators that induce the various mass terms required for the inverse seesaw mechanism.
Because these operators carry different powers of Planck suppression, the model automatically produces the necessary hierarchy $\mu \ll M$, thereby explaining both the origin and smallness of the neutrino masses without requiring fine-tuning.
Our model provides, for the first time, a well-motivated theoretical foundation for the inverse seesaw mechanism and establishes a bridge between it and the high-quality QCD axion.

Remarkably, this model also gives rise to a rich and testable phenomenology.
Our high-quality QCD axion dark matter is predicted to have a mass range of $3\times10^{-5}\,\mathrm{eV} \lesssim m_a \lesssim 5\times10^{-4}\,\mathrm{eV}$, which can be probed by upcoming haloscope experiments. 
In the neutrino sector, most of the viable parameter space falls within the sensitivity of future searches, from beam-dump experiments such as SHiP to high-energy colliders like the HL-LHC, FCC-ee, or CEPC. 
Since our model enforces a lower bound on the total mixing, a measurement of $U_\alpha^{2} < 10^{-7}$ across all flavors would provide a decisive test. 
Therefore, for $M_N \lesssim 50\,\mathrm{GeV}$, the parameter space can be probed by approved or proposed future experiments. Furthermore, it offers possible mechanisms for generating the baryon asymmetry of the Universe, through low-scale leptogenesis driven by the three pseudo-Dirac pairs.

\section{acknowledgments}
Y. G. and J. S. would like to thank Kairui Zhang for helpful discussions. 
The work of Y. G., J. S., and T. T. Y. is supported by the World Premier International Research Center Initiative (WPI), MEXT, Japan (Kavli IPMU). 
In addition, J. S. is supported by the Japan Society for the Promotion of Science (JSPS) as a part of the JSPS Postdoctoral Program (Standard) with grant number: P25018. 
S. U. has received funding from the European Union’s Horizon Europe research and innovation programme under the Marie Skłodowska-Curie Staff Exchange  grant agreement No 101086085 – ASYMMETRY. 
T. T. Y. is supported by the Natural Science Foundation of China (NSFC) under Grant No. 12175134 and MEXT KAKENHI Grants No. 24H02244. 
S. U. thanks Kavli IPMU for its hospitality during the development of this work.

\bibliographystyle{apsrev4-1}
\bibliography{Ref}

\vspace{15mm}
\end{document}